\input harvmac.tex

%%%% Useful Definitions %%%%%%%%%%%%%%%%%%%%%%%%%%%%%%%%

\def\ha{{1\over 2}}
\def\cor{\hat =}

\def\bZ{{\bf Z}}
\def\bT{{\bf T}}
\def\bCP{{\bf CP}}

%%%% Journals %%%%%%%%%%%%%%%%%%%%%%%%%%%%%%%%%%%%%%%%%%

\def\np#1#2#3{Nucl. Phys. {\bf B#1} (#2) #3}
\def\pl#1#2#3{Phys. Lett. {\bf #1B} (#2) #3}
\def\prl#1#2#3{Phys. Rev. Lett. {\bf #1} (#2) #3}

\def\physrev#1#2#3{Phys. Rev. {\bf D#1} (#2) #3}
\def\jgp#1#2#3{J. Geom. Phys. {\bf #1} (#2) #3}

%%%% References %%%%%%%%%%%%%%%%%%%%%%%%%%%%%%%%%%%%%%%%

\lref\mrdprobe{M. R. Douglas, ``Branes within Branes,''
hep-th/9512077\semi
M. R. Douglas and G. Moore, ``D-branes, Quivers, and ALE Instantons,''
hep-th/9603167\semi
M. R. Douglas, ``Gauge Fields and D-branes,'' hep-th/9604198}%
\lref\sen{A. Sen, ``F Theory and Orientifolds'', \np{475}{1996}{562},
hep-th/9605150}%
\lref\vafa{C. Vafa, ``Evidence for F-Theory'', \np{469}{1996}{403},
hep-th/9602022}%
\lref\swA{N. Seiberg and E. Witten, ``Electric-Magnetic Duality, Monopole
Condensation and Confinement in $N=2$ Supersymmetric Yang-Mills
Theory'', \np{426}{1994}{19}, hep-th/9407087}%
\lref\swB{N. Seiberg and E. Witten, ``Monopoles, Duality and Chiral
Symmetry Breaking in $N=2$ Supersymmetric QCD'', \np{431}{1994}{484},
hep-th/9408099}%
\lref\bds{T. Banks, M. R. Douglas and N. Seiberg, ``Probing F-Theory With
Branes'', \pl{387}{1996}{278}, hep-th/9605199}%
\lref\seiberg{N. Seiberg, ``IR Dynamics on Branes and Space-time
Geometry'', \pl{384}{1996}{81}, hep-th/9606017}%
\lref\seibergfive{N. Seiberg, ``Five-Dimensional SUSY Field Theories,
Nontrivial Fixed Points and String Dynamics'', hep-th/9608111}%
\lref\gms{O. J. Ganor, D. R. Morrison and N. Seiberg, ``Branes,
Calabi-Yau Spaces, and Toroidal Compactification of the N=1
Six-Dimensional $E_8$ Theory'', hep-th/9610251}%
\lref\pol{J. Polchinski, ``Dirichlet Branes and Ramond-Ramond
Charges'', \prl{75}{1995}{4724}, hep-th/9510169}
\lref\gp{E. G. Gimon and J. Polchinski, ``Consistency Conditions for
Orientifolds and D-Manifolds'',
\physrev{54}{1996}{1667}, hep-th/9601038}%
\lref\as{P. C. Argyres and A. D. Shapere, ``The Vacuum Structure of $N=2$
SuperQCD with Classical Gauge Groups'', \np{461}{1996}{437},
hep-th/9509175}%
\lref\witbound{E. Witten, ``Bound States of Strings and $p$-branes'',
\np{460}{1996}{335}, hep-th/9510135}%
\lref\dw{R. Donagi and E. Witten, ``Supersymmetric Yang-Mills Theory and
Integrable Systems'', \np{460}{1996}{299}, hep-th/9510101}%
\lref\mv{D. R. Morrison and C. Vafa, ``Compactifications of F-Theory on
Calabi-Yau Threefolds I,II'',
\np{473}{1996}{74}, hep-th/9602114;
\np{476}{1996}{437}, hep-th/9603161}%
\lref\bz{J. D. Blum and A. Zaffaroni, ``An Orientifold from F-Theory'',
\pl{387}{1996}{71}, hep-th/9607019}%
\lref\dabpark{A. Dabholkar and J. Park, ``A Note on Orientifolds and 
F-Theory'', hep-th/9607041}%
\lref\gm{R. Gopakumar and S. Mukhi, ``Orbifold and Orientifold
Compactifications of F-Theory and M Theory to Six and Four Dimensions'',
hep-th/9607057}%
\lref\micha{M. Berkooz, R. G. Leigh, J. Polchinski, J. H. Schwarz, N.
Seiberg and E. Witten, ``Anomalies, Dualities and Topology of $D=6$ $N=1$
Superstring Vacua'', \np{475}{1996}{115}, hep-th/9605184}%
\lref\witten{E. Witten, ``Some Comments on String Dynamics'', contributed
to Strings '95, hep-th/9507121}%
\lref\ori{O. J. Ganor and A. Hanany, ``Small $E_8$ Instantons and
Tensionless Non-Critical Strings'',
\np{474}{1996}{122}, hep-th/9602120}%
\lref\swC{N. Seiberg and E. Witten, ``Comments on String Dynamics in Six
Dimensions'',
\np{471}{1996}{121}, hep-th/9603003}%
\lref\phase{E. Witten, ``Phase Transitions in M Theory and F-Theory'',
\np{471}{1996}{195}, hep-th/9603150}%
\lref\swD{N. Seiberg and E. Witten, ``Gauge Dynamics and Compactification
to Three Dimensions'', hep-th/9607163}%
\lref\dual{N. Seiberg, ``Electric-Magnetic Duality in Supersymmetric
Nonabelian Gauge Theories'', \np{435}{1995}{129}, hep-th/9411149}%
\lref\ls{R. G. Leigh and M. Strassler, ``Exactly Marginal Operators and
Duality in Four Dimensional $N=1$ Supersymmetric Gauge Theory'',
\np{447}{1995}{95}, hep-th/9503121}%
\lref\bikmsv{ M. Bershadsky, K. Intriligator, S. Kachru,
D. R. Morrison, V. Sadov and C. Vafa,
``Geometric Singularities and Enhanced Gauge
Symmetries", hep-th/9605200}%
\lref\bj{M. Bershadsky and A. Johansen, ``Colliding Singularities in F
Theory and Phase Transitions'', hep-th/9610111}%
\lref\apsw{P. C. Argyres, M. R. Plesser, N. Seiberg and E. Witten,
``New $N=2$ superconformal field theories in four dimensions'',
\np{461}{1996}{71}, hep-th/9511154}%
\lref\aks{O. Aharony, S. Kachru and E. Silverstein, ``New $N=1$
superconformal field theories in four dimensions from D-brane
probes'', hep-th/9610205}%
\lref\eseries{J. Minahan and D. Nemeschansky, ``An ${\cal N}=2$
Superconformal Fixed Point
with $E_6$ Global Symmetry,'' hep-th/9608047\semi
W. Lerche and N. Warner, ``Exceptional SW Geometry from ALE
Fibrations,'' hep-th/9608183\semi
J. Minahan and D. Nemeschansky, ``Superconformal Fixed Points with $E_n$
Global Symmetry,'' hep-th/9610076}%
\lref\ori{O. J. Ganor, ``Toroidal Compactification of Heterotic 6D
Noncritical Strings Down to Four Dimensions'', hep-th/9608109}%
\lref\wittensmall{E. Witten, ``Small Instantons in String Theory'',
\np{460}{1996}{541}, hep-th/9511030}%
\lref\bhoo{J. de Boer, K. Hori, H. Ooguri and Y. Oz, ``Mirror Symmetry
in Three-Dimensional Gauge Theories, Quivers and D-branes'',
hep-th/9611063}%
\lref\sentwo{A. Sen, ``A Non-perturbative Description of the Gimon-Polchinski
Orientifold'', hep-th/9611186}%
\lref\disctor{C. Vafa and E. Witten, ``On Orbifolds With Discrete
Torsion'', \jgp{15}{1995}{189}, hep-th/9409188}%
\lref\paul{P. Aspinwall, ``Enhanced Gauge Symmetries and K3
Surfaces'', \pl{357}{1995}{329}, hep-th/9507012}%

%%%% Title Page %%%%%%%%%%%%%%%%%%%%%%%%%%%%%%%%%%%%%%%%%%%%%

% Title page

\def\Title#1#2{\nopagenumbers\abstractfont\hsize=\hstitle\rightline{#1}%
\vskip .8cm\centerline{\titlefont #2}\abstractfont\vskip .3cm\pageno=0}

\hfill{RU-96-105, TAUP-2397-96,
LMU-TPW-96-33, CERN-TH/96-337}
\Title{hep-th/9611222}
{\vbox{\centerline{Field Theory Questions for String Theory Answers}}
\footnote{$^*$}{Work supported in part by the
US-Israel Binational Science Foundation, by GIF -- the German-Israeli
Foundation for Scientific Research, and by the Israel Academy of
Science.}}

\medskip

\centerline{\bf Ofer Aharony}
\centerline{Department of Physics and Astronomy}
\centerline{Rutgers University}
\centerline{Piscataway, NJ 08855-0849}
\vglue .5cm
\centerline{{\bf Jacob Sonnenschein}
and {\bf Shimon Yankielowicz}} 
\centerline{School of Physics and Astronomy}
\centerline{Beverly and Raymond Sackler Faculty of Exact Sciences}
\centerline{Tel--Aviv University}
\centerline{Ramat--Aviv, Tel--Aviv 69978, Israel}
\vglue .5cm
\centerline{\bf Stefan Theisen}\footnote{}{
E-mail addresses: oferah@physics.rutgers.edu,
cobi@ccsg.tau.ac.il,
shimonya@ccsg.tau.ac.il, 
theisen@mppmu.mpg.de}
\centerline{Sektion Physik, University of Munich, Munich, Germany
\foot{Permanent address}}
\centerline{and}
\centerline{CERN, Theory Division, Geneva, Switzerland}

\medskip

{
We discuss the field theory of 3-brane probes in F-theory
compactifications in two configurations, generalizing the work of Sen
and of Banks, Douglas and Seiberg. One configuration involves several
parallel 3-brane probes in F-theory compactified on $\bT^4 / \bZ_2$,
while the other involves a compactification of F-theory on $\bT^6 /
\bZ_2 \times \bZ_2$ (which includes intersecting $D_4$ singularities). 
In both cases string theory
provides simple pictures of the spacetime theory,
whose implications for the three-brane world-volume
theories are discussed.
In the second case  the field theory on the probe is 
an unusual $N=1$ superconformal theory, with exact
electric-magnetic duality. Several open questions remain concerning
the description of this theory.
}

\noindent

\Date{11/96}

% Paper
%%%% Paper %%%%%%%%%%%%%%%%%%%%%%%%%%%%%%%%%%%%%%%%%%%%%%%%%%

%\leftline{\bf{Preliminary Draft:\number\day/\number\month/\number\yearltd\ }}

\newsec{Introduction}

The study of brane probes in string theory vacua enables us to connect,
in a relatively simple way, results in field theory with results in
string theory. The first such example involving a 
four dimensional field theory
was described in \refs{\sen,\bds}.
Sen showed \sen\ that an orbifold limit of F-theory \vafa\ on K3
was T-dual to type I
compactified on $\bT^2$, and that F-theory provided a consistent description
of the smoothing of the orientifold points when moving away from
the orientifold limit.
The mathematical description of the
region (both in space-time and in moduli space) around an orientifold point
was found to be exactly the same as Seiberg and Witten's description of the
quantum-corrected moduli space of the $SU(2)$ $N=2$ gauge theory with
$N_f=4$ \refs{\swA,\swB}.
Banks, Douglas and Seiberg \bds\ later showed that this similarity was not a
coincidence. It arises just by looking at the effective field theory on the
worldvolume of a 3-brane moving in this vacuum, and demanding consistency
between the quantum corrections in this field theory and in the spacetime
theory (whose fields couple to the 3-brane worldvolume). This sort of
relation (first discussed in \mrdprobe) should
presumably hold for any D-brane
moving in a string theory
vacuum, as long as the probe does not modify the vacuum in a
substantial way. It provides a general relation between string theory,
and its low energy effective spacetime theories, and
the worldvolume field theories of D-branes.
The D-branes thus
may serve as probes which enable us to learn more about non-perturbative
aspects of string theory (and of field theory).
When the resolution
of the singularities in string theory is known, we can use it to learn about
the resolution of singularities in field theory, and vice versa.
This approach has since been generalized to several other cases (for
instance, in \refs{\seiberg,\seibergfive,\aks,\gms}).

In this note we wish to analyze two generalizations of this approach.
First, in section 2, we discuss the field theory of multiple parallel
D-branes as probes, in the
same string theory vacuum considered in \refs{\sen,\bds}.
In section 3 we analyze a generalization to six dimensions,
discussing F-theory on $\bT^6/\bZ_2\times \bZ_2$ (or, more precisely,
an orientifold which differs from this by discrete torsion). In both
cases the string theory (F-theory) ``answer'' is trivially known, and we
are looking for the classical field theory on the 3-brane which reproduces
this
``answer'' when computing its quantum moduli space. In the first case this
field theory turns out to be a simple $N=2$ field theory, for which it is
easy to see also directly in field theory that the moduli space decomposes
into several copies of the Seiberg-Witten moduli space. In the second case
we find an $N=1$ superconformal field theory which does not seem to be
known, and which has many unusual properties, including an exact
electric-magnetic duality. We compute the field content of
this theory, but we do not know how to write down an explicit
superpotential to describe it. In section 4 we give a summary of
our results and of the remaining open questions. In an appendix we
describe explicitly our calculations of the field content of the
3-brane probe theory near an orientifold point in the string theory
discussed in section 3.
As this paper was being completed, a preprint by Sen \sentwo\ appeared, in
which he treats a different orientifold, namely a $T$-dual version of
the Gimon-Polchinski \gp\ model. The analogous analysis 
for this model, which leads to a different space-time and also 
world-volume theory, can be easily carried through and one finds that the
problems we have encountered with our original model are absent there. 
We have included a brief discussion of this model, along the lines 
exposed in section three, at the end of section four. Further 
details of this model, in particular the space-time aspects and its
non-perturbative description, are found in \sentwo.

\newsec{Multiple 3-branes as probes in 8 dimensions}

In this section we analyze a trivial generalization of the approach of
Banks, Douglas and Seiberg \bds, which involves using $k$ 3-branes as probes
instead of a single 3-brane. Of course, since there are no forces between
parallel 3-branes \pol\
(and we will use only parallel 3-branes), we expect the
result to be the same as the result found by Sen. Each 3-brane should see one
region of the
same vacuum of the string theory, according to its spacetime position, and
the moduli space should be the product of the moduli spaces of the different
3-branes (up to an $S_k$ ``Weyl group" which permutes the identical 3-branes).
This expectation will be confirmed by the field theory analysis of the
worldvolume theory on the 3-branes, as we will now show.

As in \bds, we will begin with the
description
of this theory as type I on $\bT^2$. Instead of looking at a single 5-brane
wrapped around $\bT^2$, we will now take $k$ 5-branes wrapped around
$\bT^2$.
When the 5-branes are at the same position, the field theory describing
them is well-known (see, for example, \wittensmall).
There is
an $Sp(2k)$ gauge group\foot{In this paper we denote by $Sp(2k)$ the
symplectic group whose fundamental representation is of dimension $2k$.},
with a hypermultiplet in the antisymmetric
representation (which decomposes into an irreducible $\bf k(2k-1)-1$ and
a singlet). This is an $N=1$ theory in 5+1 dimensions, which becomes an
$N=2$ theory in 3+1 dimensions when we reduce it on the torus. The Wilson
loops of the gauge field around the two compact dimensions become
expectation values of the 2 scalar fields of the $N=2$ vector multiplets in
four dimensions. The scalars in the antisymmetric representation
hypermultiplet describe the motion of the 3-branes in the four non-compact
transverse dimensions. In the 10D type I theory there are $32$ additional
half-hypermultiplets in the $\bf 2k$ representation of the $Sp(2k)$
gauge group (and in the fundamental representation of the $SO(32)$
spacetime gauge symmetry). As discussed in \bds, due to an interplay
between the $SO(32)$ Wilson loops and the $Sp(2k)$ Wilson loops, four
of these hypermultiplets are light when the probe is near an
orientifold point. In the F-theory these arise
from strings connecting the 3-brane probe with the four 7-branes which sit
at each orientifold point. As in \sen, we take the torus $\bT^2$ to be
very large and analyze only the region near a single orientifold point, when
all fields corresponding to the other points are very massive.

To summarize, the worldvolume field theory corresponding to $k$
parallel 3-branes
moving in this vacuum is an $Sp(2k)$ $N=2$ gauge theory, with four
fundamental hypermultiplets and an additional hypermultiplet in the
antisymmetric representation. The beta function of this theory is zero, as
required by the string theory picture which has no scale (when the
3-brane is exactly at the orientifold point).
The theory has an $SO(8)$ global symmetry acting on the ``quark''
hypermultiplets, which
corresponds to the $SO(8)$ gauge symmetry in spacetime. The
vector multiplet in the 7+1 dimensional spacetime includes also a
complex scalar field in the antisymmetric
(adjoint) $\bf 28$ representation of $SO(8)$. In the
worldvolume field theory, this
corresponds to a mass parameter for the quarks in the $\bf 2k$ representation.
In the F-theory picture these mass
parameters correspond to the motion of the 7-branes in the 2 compact
directions. These are the only scalar fields in spacetime, so they are
the only parameters we have in our worldvolume theory.
In the field theory one could also add a mass term for the
antisymmetric hypermultiplet, which would prevent the 3-branes from
moving in the four non-compact directions.
It would be very interesting to see whether such a mass deformation can be
accounted for in string theory (so far no string theory realization of
such a term is known).

Next, we should analyze the quantum moduli space of this field theory.
As in \bds, we will analyze only the
Coulomb phase of the theory (the Higgs phase should presumably be
related to SO(8) instantons in space-time). In this phase,
the scalar field in the vector
multiplet has an arbitrary expectation value, which may be brought (by
gauge transformations) to a diagonal matrix of the form
$diag(a_1,a_2,\cdots,a_k,-a_1,-a_2,\cdots,-a_k)$, where the $a_i$ are
arbitrary complex numbers. This VEV breaks the $Sp(2k)$ symmetry to $U(1)^k$.
The eigenvalues $a_i$ correspond (when squared) to the position
of the $k$ 3-branes in the 2 compact coordinates, measured from the
orientifold point. When we are at an arbitrary point in the Coulomb phase,
the superpotential (whose form is dictated by $N=2$ supersymmetry) does not
allow an expectation value for the hypermultiplets in the fundamental
representation. However, the flat directions corresponding to the
hypermultiplet in the antisymmetric representation are not all lifted by
the adjoint VEV, and a moduli space of $k$ dimensions remains flat
(one of these is the singlet component of the antisymmetric
representation, while the other $k-1$ are part of the $\bf k(2k-1)-1$
representation). This
corresponds to a VEV for the antisymmetric matrix of the form
\eqn\antivev{\pmatrix{0 & 0 & \cdots & 0 & b_1 & 0 & \cdots & 0 \cr
                      0 & 0 & \cdots & 0 & 0 & b_2 & \cdots & 0 \cr
         \vdots & \vdots & \ddots & \vdots & \vdots & \vdots & \vdots
& \vdots \cr
                      0 & 0 & \cdots & 0 & 0 & 0 & \cdots & b_k \cr
                      -b_1 & 0 & \cdots & 0 & 0 & 0 & \cdots & 0 \cr
                      0 & -b_2 & \cdots & 0 & 0 & 0 & \cdots & 0 \cr
     \vdots & \vdots & \vdots & \vdots & \vdots & \vdots & \ddots &
\vdots \cr
                     0 & 0 & \cdots & -b_k & 0 & 0 & \cdots & 0 \cr}.}
Each of the four scalars in the antisymmetric hypermultiplet has an
independent VEV, and they all commute so that all of them can be brought to
the form \antivev\ simultaneously. The simplest way to consider this is to
treat the $b_i$ above as quaternionic numbers.
In the absence of the adjoint VEV, the VEV \antivev\
breaks the $Sp(2k)$ symmetry to
$SU(2)^k$. The $b_i$ correspond to the position of the 3-branes in the
non-compact transverse dimensions ($x_4,x_5,x_6$ and $x_7$).
However, in $N=2$ supersymmetric theories, the moduli space
decomposes into the product of the vector multiplet moduli space and the
hypermultiplet moduli space. Thus, the vector multiplet moduli space cannot
depend on the scalars $b_i$ (which can get VEVs at arbitrary points in the
Coulomb phase). In particular, the vector multiplet moduli space remains
the same if we take all the $b_i$ to be very large, flowing to an $SU(2)^k$
gauge theory. In this limit nothing (but singlets) remains of the
antisymmetric hypermultiplet, and we remain with $k$ copies of the
Seiberg-Witten theory of $SU(2)$ gauge group with $N_f=4$. Globally we should
divide this by a discrete $S_k$ symmetry (which is a subgroup of the Weyl
group of $Sp(2k)$), since all copies (3-branes) are identical.
All copies have the same
masses for the quark hypermultiplets (corresponding to the 7-brane
positions in the F-theory description), but have independent values of the
adjoint VEVs (which are the $a_i$ above), corresponding to the position of
the 3-brane in the compact space. Thus, we have recovered the same picture
we expected to find from the string theory point of view, since there are no
interactions between the 3-branes. At special points in the moduli space we
have enhanced global symmetries, corresponding to enhanced gauge symmetries
in spacetime. The discussion of these is exactly analogous to the
discussion in \refs{\sen,\bds}.

Now that we understand the moduli space of this theory, we can try to learn
from it about other field theories we can flow to. First, let us look at
the case of $k=2$, where the gauge group is $Sp(4) \sim SO(5)$. From the
$SO(5)$ point of view, we have one fundamental hypermultiplet and four
spinors (the singlet decouples)
and (as discussed above) the moduli space decomposes into two
copies of an $SU(2)$ moduli space. Since we can give arbitrary masses to
the spinors, the same should also be true for the $SO(5)$ gauge theory with
a single fundamental hypermultiplet, which was analyzed in the past.
However, by an argument analogous to the one we used above, it is trivial
to see that the Coulomb phase of an $SO(2r+1)$ gauge theory with $N_f$
flavors (one of which is massless) is generally the same as the Coulomb
phase of an $SO(2r)$ gauge theory with $N_f-1$ flavors. This is because we
can give an arbitrary VEV to one fundamental hypermultiplet without
disturbing the Coulomb phase, and it is corroborated by the explicit form
of the curves (as written, for instance, in \as).

Another theory we can flow to is the $U(k)$ $N=4$ theory, which we can
reach by taking the 3-branes to be together but
far away from the orientifold point
\witbound.
In the gauge theory description, we reach this limit by taking
the modulus $u \to
\infty$ in all of the $SU(2)$ curves,
and we arrive at $k$ copies of the (free)
$U(1)$ gauge theory (up to the $S_k$ identification). Indeed, this is
the correct description of the $U(k)$ $N=4$ theory, as analyzed by Donagi
and Witten \dw. We can also show this, as above, by decomposing the $N=4$
vector multiplet into an $N=2$ vector multiplet and hypermultiplet, and
giving large VEVs to the hypermultiplet (which should not affect the vector
moduli space), breaking the gauge symmetry to $U(1)^k$.

To summarize, multiple $p$-branes may also be used to learn from field
theory about string theory or vice versa. However, in the Coulomb
phase of the field theory they
just lead to $k$ copies (up to global identifications) of the Coulomb
phase of the field theory
found for a single $p$-brane. We can show, using the string theory, that
the moduli space of these field theories indeed decomposes in this way, but
so far in all the examples it is easy to see this also directly from field
theory arguments.
The superconformal theory at the
origin of the moduli space is, however, not trivially related to the
theory found for a single 3-brane\foot{We thank N. Seiberg for
stressing this point.}. For instance, we can turn on a mass for the
antisymmetric hypermultiplet in this theory and flow to an $N=2$ $Sp(2k)$
gauge theory with $N_f=4$, which behaves differently for different
values of $k$. Thus, the analysis of multiple brane probes may lead to
the discovery of new superconformal field theories, though their
Coulomb phases would always be simply related to those of the single
probe theory.

\newsec{$N=1$ 4d SCFTs from 3-branes at intersecting singularities}

In F-theory compactifications to 8 dimensions, the 3-brane probe has
$N=2$ supersymmetry. Interesting $N=2$ SCFTs arise when the 3-brane is
brought to a singularity of the compactification manifold. Non-trivial
conformal field theories arise when the singularity is of type
$A_0,A_1,A_2,D_4,E_6,E_7$ or $E_8$. The first four correspond to known
$N=2$ SCFTs \apsw\ while the other $N=2$ SCFTs were recently discussed
from various points of view in \refs{\eseries,\ori,\gms}.
Going down to 6 dimensions,
there are many more possibilities for singularities in the manifold,
and generically the 3-brane field theory has only $N=1$
supersymmetry. In \aks, the behavior of a 3-brane probe near an ADE
singularity fibered over the additional two compact dimensions
\bikmsv\ was
analyzed, leading to several new $N=1$ SCFTs. In this section we
discuss the behavior of a 3-brane probe near the intersection of two
singularities, which also corresponds to $N=1$ SCFTs.

The only singularity which can occur at a weak (but non-zero) value of
the coupling is a $D_4$ singularity \swB,
so this is the only case where we
definitely expect to have a lagrangian description of the probe
theory. As in \sen, let us look for compactifications of F
theory on CY manifolds which have a constant value of $\tau$ (which we
can take to correspond to weak coupling).
The simplest case in which this occurs is
the elliptic fiber over a base $\bCP^1\times \bCP^1$ \mv. In general
such a fibration is described by an equation of the form
\eqn\fibration{y^2 = x^3 + f(z_1,z_2) x + g(z_1,z_2),}
where $z_1$ and $z_2$ label the two $\bCP^1$s of the base.
The torus of the fiber degenerates where
the discriminant $\Delta=4f^3+27g^2$ vanishes, and in F-theory the
solutions to $\Delta=0$ are the locations of type IIB 7-branes in the
compact dimensions,
and $\tau$ has a non-trivial
monodromy when going around them. $j(\tau(z_1,z_2))$ is then
proportional to
$f^3/\Delta$, and we want this to be constant. The simplest
choice (though not the most general solution) is of the form
\eqn\choice{f(z_1,z_2) = \alpha \phi_1(z_1)^2 \phi_2(z_2)^2; \qquad
g(z_1,z_2) = \phi_1(z_1)^3 \phi_2(z_2)^3,}
where $\phi_1$ and $\phi_2$ are general polynomials of degree four :
\eqn\phis{\phi_1(z_1) = \prod_{i=1}^4 (z_1 - z^{(i)}_1); \qquad \phi_2(z_2)
= \prod_{i=1}^4 (z_2 - z^{(i)}_2).}
It is clear that at this point in moduli space $\tau$ has a constant value
depending on $\alpha$, as in \sen, and that we have $D_4$ singularities at
$z_1=z_1^{(i)}$ and at $z_2=z_2^{(i)}$.

In general the spacetime field theory at an intersection of
two $D_4$ singularities is not well understood \refs{\bikmsv,\bj}. However,
as in Sen's case \sen, we can try to
interpret this point in moduli space as an
orientifold of the type IIB theory, and since it can be weakly coupled
we can analyze the spacetime field theory by perturbative string
theory methods. Around each of the points $z_1=z^{(i)}_1$ (for
constant $z_2$) we have an $SL(2,\bZ)$ monodromy of
\eqn\monodromy{\pmatrix{-1 & 0 \cr 0 & -1 \cr}.}
Locally,
we can interpret this as an orientifold of the type IIB theory by $(-1)^{F_L}
\cdot \Omega$, as discussed in \sen, which should be accompanied by four
7-branes to cancel the RR charge. Each point $z_1=z^{(i)}_1$
on the first $\bCP^1$ factor carries a
deficit angle of $\pi$, all four of them together
deforming the $\bCP^1$ to $\bT^2/\bZ_2$. The same is true at the points
$z_2=z^{(i)}_2$, and altogether 
it seems that we can write this theory as the type IIB
theory on $\bT^4$,
divided by a product of two $\bZ_2$ symmetries of the form found by
Sen \sen, one of which inverts $z_1$ while the other inverts $z_2$. 
This
corresponds to the orientifold of F-theory on $\bT^6$ by $(\bZ_2)^2$,
discussed in \refs{\bz,\dabpark,\gm}, which
can be deformed into the $(51,3)$ CY manifold (by changing the values
of tensor-multiplet scalars).

However, we should be careful in identifying this orientifold
compactification directly with the F-theory compactification described
by \fibration-\phis. Generally, in F-theory compactifications, the
2-form tensor fields of type IIB are assumed to vanish. This is, of
course, consistent with the $SL(2,\bZ)$ monodromies that these fields
have around singularities of the elliptic fibration. However, at the
intersection of two $D_4$ singularities that we have been discussing,
there is the possibility of blowing up the intersection
point to get an additional 2-cycle \refs{\bikmsv,\bj}. 
When there are vanishing 2-cycles, the possibility arises of
having discrete 2-form fields concentrated on the vanishing
2-cycle, a phenomenon which is related in string theory to discrete
torsion \disctor. In the naive F-theory compactification, we assume
that these fields are zero, and then one can wrap a 3-brane around the
vanishing 2-cycle, giving rise to tensionless strings living on the
intersection of the $D_4$ singularities. Our orientifold construction,
on the other hand, involves a well-defined weakly coupled conformal
field theory, whose low-energy spectrum is well-defined and does not
include such tensionless strings\foot{We thank E. Witten for
emphasizing this point.}. Thus, we conclude that the
orientifold differs from the F-theory construction by a discrete
2-form field, which prevents the 3-brane from wrapping around the
vanishing 2-cycle (a similar phenomenon involving wrapped 2-branes
was described in
\paul). Discrete $\bZ_2$ symmetries force the value of both 2-form
tensor fields (integrated over the vanishing 2-cycle) to be either $0$
or $1/2$ (modulo $1$). 
A non-zero value for either (or both) of these fields breaks
the $SL(2,\bZ)$ symmetry to a discrete $\Gamma(2)$ subgroup. For
instance, if both fields equal $1/2$, this subgroup is generated by
$S$ and by $T^2$ (where $S$ and $T$ are the standard $SL(2,\bZ)$
generators; note that in the cases we are discussing the string
coupling $\tau$ equals $\tau = {{8\pi i}\over g^2} + {\theta \over
\pi}$ in the field theory, so only $T^2$ is naively guaranteed to be a
symmetry, as in \swB). The other cases are related to this by $SL(2,\bZ)$
transformations. In the remainder of this section, we will discuss a
3-brane probe in the orientifold background, which is thus expected to
describe an $N=1$ SCFT with a $\Gamma(2)$ electric-magnetic duality
symmetry. At the end of the section we will comment on the relation
between this theory and the theory of a probe in the F-theory
background with no 2-form fields turned on.

The spacetime theory corresponding to the orientifold is an $N=1$
supersymmetric theory, whose field content was computed in
\refs{\bz,\dabpark,\gm}. The untwisted closed string sectors give rise
to a supergravity multiplet, to one tensor multiplet and to four
hypermultiplets. The open string sectors give an $SO(8)$ gauge group
for each group of 7-branes (i.e. each fixed point of one of the
$\bZ_2$'s), for a total gauge symmetry of $SO(8)^8$. No massless states
arise from strings between the different groups of 7-branes, but the
twisted closed string sector gives rise to a tensor multiplet at each
intersection of fixed points (i.e. at each fixed point of both
$\bZ_2$'s), accounting for a total of 16 additional tensor multiplets. The
spacetime theory satisfies all of the anomaly cancelation
requirements. 
At each intersection of $D_4$ singularities
there is a possibility of blowing up a point to get an additional 2-cycle
(by turning on the scalar in the corresponding tensor multiplet). 
After all these blow-ups we get F-theory on a smooth $(51,3)$
Calabi-Yau manifold.
Note that the spacetime theory we found (using the orientifold
construction) is not the same as the theory we supposedly started
with, which was F-theory compactified on an elliptic fibration over
$\bCP^1 \times \bCP^1$. It may be possible to reach that theory by
performing phase transitions which turn the 16 additional tensor
multiplets we have into hypermultiplets, but we cannot discuss this
directly in the orientifold construction.

We would like to analyze what happens to a 3-brane probe as it moves
around in this string theory vacuum. Near each fixed line (i.e. a
fixed point of one of the two $\bZ_2$'s), the spacetime theory is the same
as the one analyzed in \refs{\sen,\bds} and in the previous
section. Thus, the theory on the probe should be just an $N=2$ $SU(2)$
gauge theory with 4 quark hypermultiplets. Things get more interesting
if we move the probe to the intersection of two of the fixed lines
(which we will take to be at $z_1=z_2=0$). We expect the probe theory
there to have only $N=1$ supersymmetry, since the 7-branes intersect
transversely at this point and each breaks a different half of the
worldvolume $N=4$ supersymmetry. The probe theory should have two
deformations, corresponding to turning on $z_1$ or $z_2$, which
should cause it to flow to the $N=2$ $SU(2)$ theory.

In general, it is difficult to analyze the worldvolume theory of a
3-brane probe in F-theory vacua when it is adjacent to 7-branes of
different types (which are not just D-7-branes). However, since
in this case
we have an orientifold description of this vacuum, we can use it
to compute the states living on the 3-brane. In this way we can
compute the states on the 3-brane arising from fundamental strings,
which will lead to electrically charged states on the 3-brane. Since
the theory on the probe is manifestly invariant under a $\Gamma(2)$
electric-magnetic duality symmetry, we expect that there will also be
$(p,q)$ strings which give rise to $(p,q)$ dyonic states in the
worldvolume theory. However, since our theory has only $N=1$
supersymmetry and no BPS formulas, it is not obvious that these states
should remain stable also at weak electric coupling (unlike the $N=2$
case).

Let us begin by computing the fields
corresponding to strings which stretch out from a 3-brane and fold back to the
same brane. The 3-brane has 3 images under $\bZ_2 \times
\bZ_2$, so that every open string state is enhanced to a $4\times 4$
matrix (as in \gp), on which we should impose the orientifold restrictions.
The $\gamma$ matrices (in the notation of \gp)
corresponding to each of the generators in the $\bZ_2 \times \bZ_2$
may be chosen to be
\eqn\gammaone{
\gamma_{\Omega_1} = \pmatrix{ 0 & i & 0 & 0 \cr -i & 0 & 0 & 0 \cr
0 & 0 & 0 & i \cr 0 & 0 & -i & 0 \cr}  \quad
\gamma_{\Omega_2} = \pmatrix{0 & 0 & i & 0 \cr 0 & 0 & 0 & i \cr -i &
0 & 0 & 0 \cr 0 & -i & 0 & 0 \cr}}
and the orbifold matrix is then necessarily
\eqn\gammatwo{
\gamma_T = \gamma_{\Omega_1} \gamma_{\Omega_2} = \pmatrix{0 & 0 & 0
& -1 \cr 0 & 0 & 1 & 0 \cr 0 & 1 & 0 & 0 \cr -1 & 0 & 0 & 0 \cr}.}
These matrices are hermitian and unitary, and they correctly
reflect the two $\bZ_2$ actions on the compact coordinates $(z_1,z_2)$.

The wave-function matrices of states, $M_{ij}$, must then satisfy \gp\
conditions of the type
\eqn\mishgam{\eqalign{M &= \pm \gamma_{\Omega_1} M^T
\gamma_{\Omega_1}^{-1} \cr
M &= \pm \gamma_{\Omega_2} M^T \gamma_{\Omega_2}^{-1} \cr
M &= \pm \gamma_T M \gamma_T^{-1} \cr}}
where the signs are determined by the transformation properties of the
relevant state.
For the gauge fields they are $-,-,+$, for the
chiral superfield $X_{6,7}$
they are $-,+,-$, for  $X_{8,9}$ they are $+,-,-$ and for
$X_{4,5}$ they are $+,+,+$.

Performing this computation, details of which can be found in the appendix,
we find that the gauge fields on the
3-brane worldvolume give rise to 6 states, which
generate an $SU(2) \times SU(2)$ algebra. The fields $X_{6,7}$ and
$X_{8,9}$ each give rise to 4 states, which are in the $\bf(2,2)$
representation of the gauge group, and which are in $N=1$ chiral
multiplets which we will denote by $A$ and $B$. As in \bds, we will
identify the gauge singlet
$A^2$ with the $z_1$ coordinate of the 3-brane, and $B^2$
with the $z_2$ coordinate of the 3-brane. The fields
$X_{4,5}$ give rise to two gauge-singlet chiral multiplets (which we
will denote $S_1$ and $S_2$).

There are several interesting things to note about this particle
spectrum. First, the spectrum we found
is manifestly not $N=2$ supersymmetric (since
there is no chiral multiplet in the adjoint representation). Next, we
expect that giving a VEV to $A$ ($B$) corresponds to moving in the $z_1$
($z_2$) direction, which should lead us to the $N=2$ theory. And indeed,
giving a VEV to $A$ ($B$) breaks the $SU(2)\times SU(2)$ gauge group
to an $SU(2)$ subgroup,
three components of $A$ ($B$) are
swallowed by the Higgs mechanism, and we remain with an adjoint chiral
multiplet (coming from $B$ ($A$)) and additional singlets, as
expected. It is less clear why an additional scalar appears from
$X_{4,5}$ -- from the analysis it is obvious that this additional
scalar is  massless only at the intersection point. The
natural interpretation of this is that at the intersection point
the 3-brane can split into two half-3-branes which can move
independently (this is reasonable since at orientifold points
the minimum quantum of RR charge is usually halved).

Next, we consider the 3-7 strings, namely, strings
that  stretch between the 3-brane and a 7-brane, giving rise to fields
on the 3-brane world-volume field theory.
The  $\gamma_{\Omega}$ matrices for the
7-branes were computed in \refs{\bz,\dabpark}, and they are
all proportional to the identity matrix.
As shown in the appendix, one group
of 7-branes (i.e. 7-branes located at a particular value of $z_1$)
gives rise to 8 chiral multiplets in the $\bf(2,1)$ representation of
the gauge group, which we will denote by $Q^i$ ($i=1,\cdots,8$),
while the other group (i.e. 7-branes located at a
particular value of $z_2$) gives rise to 8 chiral multiplets in the
$\bf(1,2)$ representation; those we will denote by $q^i$ ($i=1,\cdots,8$).

Let us now analyze in detail the worldvolume gauge theory at the
orientifold point. Recall that we expect this theory to be an $N=1$
superconformal field theory, which should exist for any value of the
gauge coupling (at least, any value of the $U(1)$ gauge coupling
should be possible when we turn on small values of $z_1$ and
$z_2$). Thus, we expect this theory to have a fixed line in the space
of couplings, which passes through weak coupling.

The gauge group we found is $SU(2)\times SU(2)$. There are
two chiral superfields in the $\bf(2,2)$
representation ($A$ and $B$), 8 chiral
superfields $Q$ in the $\bf(2,1)$ representation and 8 more chiral superfields
$q$ in the $\bf(1,2)$ representation
(this matter content can be
reinterpreted in terms of an $SO(4)$ gauge symmetry
with 2 vectors and 8 spinors).
In the terminology of $N=1$, the theory has $N_f=6$
for each $SU(2)$ group factor and therefore, the one-loop beta
function of the gauge coupling vanishes (as we would expect for a
theory which has a fixed line passing through weak coupling).

Next, we should compute the superpotential of this theory. In
principle, it is possible to compute this superpotential from the
string theory analysis, but we have not performed this
computation. Since we have only $N=1$ supersymmetry, the
superpotential in general receives non-perturbative quantum corrections.
However, we will try to guess what the superpotential should be
by demanding that it
reproduces our expectations of this theory. There are three
main constraints on the superpotential :

1) The theory should have (at least) an $SO(8) \times SO(8)$ global
symmetry (since this is the spacetime gauge symmetry localized at the
intersection point).

2) The theory should flow to the $N=2$, $SU(2)$, $N_f=4$ theory upon
giving a VEV to $A$ or to $B$.

3) The theory should have a fixed line passing through weak coupling.

\noindent
Note that at the orientifold point the 3-brane field theory is
conformal, and terms of degree higher than 3 in the superpotential may
also be important (as in \refs{\ls,\aks}).

First, it is easy to see what the superpotential should be for the
fields arising from $X_{4,5}$. The field $S_1$ corresponding to the
location of the 3-brane in these coordinates should be decoupled,
while the field $S_2$ (corresponding to splitting the 3-brane at the
orientifold point) should be massless only when $A$ and $B$ are both
zero. Thus, it is natural to guess a superpotential of the
form\foot{We thank A. Sen for this simplification to our
original suggestion.} $W = S_2 A B$.
With this superpotential, $A$ and $B$ become massive once
$S_2$ is turned on, as expected since when the 3-brane has split we
cannot move it away from the orientifold point.

The appearance of such a term in the superpotential is consistent also
with the dimension we expect to find for the Coulomb branch of the
theory (the phrase ``Coulomb branch'' here refers to a phase in which
$A$ and $B$ may obtain VEVs, but the quarks do not).
We know that turning on $A^2$ and $B^2$ should
correspond to flat directions of the field theory (when
$S_2=0$), since the 3-brane should be free to move in the $z_1$ and
$z_2$ directions, but these should be the only flat directions in the 
Coulomb phase. Thus, it should not be
possible to turn on a VEV for the gauge singlet $AB$, and 
this is exactly the effect of the superpotential $W = S_2 A B$.
This superpotential is also consistent with
the flow to the $N=2$ theory upon turning on $A$ or $B$. If we give a
VEV to $A$, three of its components are swallowed by the Higgs
mechanism, and the other remains massless and parametrizes the flat
direction corresponding to the flow (it is the $N=2$ partner of
$S_1$). Three of the components of $B$
remain massless and become an adjoint field $X$ of the remaining $SU(2)$,
but the remaining component, as well as $S_2$, should become massive 
(since we have no corresponding fields in the $N=2$ theory), and this 
is indeed the case if such a superpotential exists.

Finally, we should analyze which terms involving the quark field
appear in the superpotential. The requirement of an $SO(8)\times
SO(8)$ global symmetry (acting on the quarks in the obvious way), and
of $SU(2)\times SU(2)$ gauge invariance, severely limits the possible
terms that may appear. In fact, the only possible terms of degree four
or less are $W = Q^i A B Q^i$ (which can be multiplied by some function
$H(A^2,B^2)$) and $W = q^i A B q^i$ (which would then
be multiplied by
$H(B^2,A^2)$). When we give a VEV to (say) $A$, these terms flow to
the expected $N=2$ superpotential $q^i X q^i$, up to some function of
$A^2$ and of $B^2 \sim \tr(X^2)$. This function should equal 1 (or
flow to 1) when $B=0$, so that for large $A$ and small $B$,
$H(B^2,A^2)$ should behave like $1/\sqrt{B^2}$.

The superpotential terms we have written fulfill several  of our
expectations, but there are two questions which we have not yet
addressed that do not seem to be answered by these
superpotentials. First, we expect the quarks $q^i$ ($Q^i$) to become
massive when we give a VEV to $A$ ($B$), so that we will flow to a
theory with only 8 quark doublets. Note that since we are now in an
$N=1$ theory, there is no BPS mass formula from which we can compute
the exact mass of these fields, but it seems obvious that they should
become massive for non-zero $A$ ($B$). Unfortunately, it is not
possible to write down an appropriate term in the superpotential that
will preserve both the gauge symmetry and the global symmetry and that
will give such a mass to the fields. For non-zero $A$ and $B$, we can
write a superpotential of the form
\eqn\maybepot{W = Q^i A B Q^i / \sqrt{A^2} + q^i A B q^i / \sqrt{B^2}}
that flows to the correct $N=2$ superpotential and gives the expected
masses to the quarks. However, this superpotential is not well-defined
in the limit of $A\to 0$ (or $B\to 0$), and there do not seem to be
any other fields in the worldvolume theory which would enable us to
smooth this singularity.

A second problem is that we expect the theory
to have a fixed line passing through weak coupling.
To see the implications we follow, e.g., \ls, where it is shown that for
a theory with no scale dependence we need to require that the scaling
coefficients
\eqn\ag{\eqalign{
A_g&=b_1+\sum_i T(R_i)\gamma_i\cr
A_{h_n}&=(n-3)+{1\over2}\sum_k\gamma_k}}
vanish. Here, the sum in the scaling coefficient $A_g$ for the gauge couplings
is over all fields in the theory, whereas
the sum in $A_{h_n}$ is over all fields (with multiplicities)
appearing in a degree $n$ term in the superpotential; $h_n$ is the
corresponding coupling constant and $b_1$ is the one-loop
beta-function coefficient
$b_1=3C_2(G) -\sum_i T(R_i)$, which vanishes in our case. Having a fixed
manifold of dimension higher than zero requires that the scaling
coefficients are not all independent. In addition,
having a fixed line through
weak coupling, where the anomalous dimensions are expected to vanish,
requires $d_n=3$. However, in our case it is impossible to write
simple polynomial potentials of degree 3, but only superpotentials of
the form \maybepot, whose interpretation is not clear. Note that if we
use the superpotential \maybepot, the equations in \ag\ are linearly
dependent, so a fixed line (passing through weak coupling) is expected
to exist, but we do not understand how such a superpotential may arise.

As in \aks, we can compute the dimensions of some of the operators
of the fixed-point theory from the string theory metric. The elliptic
curve describing the theory near the fixed point looks like $y^2 \sim
x^3 + z_1^2 z_2^2 x + z_1^3 z_2^3$, so that $[z_1]+[z_2]=[x]$, and the
equation $[z_1]+[z_2]+[x]-[y]=2$ \aks\ leads to $[z_1]+[z_2]=4$, so
that $[A]+[B]=2$ and it is natural to assume that $[A]=[B]=1$, so that
both fields have no anomalous dimension. This is consistent with
having a fixed line that goes through weak coupling along which the
dimensions do not change, as was the case for the $N=2$ theory (and
for the $N=1$ theory discussed in \aks). If $A$ and $B$ have no
anomalous dimensions, then the requirement of the vanishing of the
beta function \ag\ leads to $Q$ and $q$ having no anomalous dimensions
either. Then, it is clear that, even if we do not demand that the
fixed line passes through weak coupling, only dimension 3 superpotentials
could be relevant at the fixed point. Again, this is consistent with
using the superpotential \maybepot\ whose interpretation is not clear,
but not, as far as we could see, with any other superpotentials.

To summarize, we have computed by a perturbative string theory
analysis the spectrum of the 3-brane probe worldvolume theory near an
intersection of two $D_4$ singularities in a particular global context
(i.e. an orientifold which is equivalent, up to discrete torsion, to
F-theory compactified on $\bT^6 / (\bZ_2 \times \bZ_2)$). The
field content we have found is consistent with our expectations, but
we have not been able to write down a superpotential at the
intersection point that will reproduce all of our expectations
(although we can write a superpotential away from the orientifold
points that does seem to be consistent). It seems that the field
theory at the fixed point cannot be described by a simple
(i.e. polynomial) superpotential. This could be related to the fact
that the theory has massless magnetic and dyonic degrees of freedom as
well as electric. However, in the $N=2$ case that did not prevent a
simple superpotential description from existing (of course, the
supersymmetry constraints were much stronger in that case). Another
possibility, which seems less likely,
is that in the case we are discussing the fields
corresponding to the other orientifold points no longer decouple, and
must all be included in the analysis.
A third
possibility is that our problems are related to the existence of a tensor
multiplet in the spacetime theory the probe
couples to at the intersection point. In principle, we should be able
to include in our field theory the coupling to the spacetime tensor
multiplet as well. The scalar in this multiplet should correspond to a
real scalar parameter of the 3-brane field theory. Turning on this
scalar corresponds to blowing the point $z_1=z_2=0$ into a 2-cycle,
after which the
two $D_4$ singularities no longer meet.
In the field theory the moduli
space should change in a similar way, and the origin of moduli space
should be blown up.
There should no longer be a point of unbroken
$SU(2)\times SU(2)$, but only non-intersecting lines of unbroken
$SU(2)$. Obviously, the appropriate terms are in the K\"ahler
potential and not in the superpotential, and we have not analyzed the
exact form that they should take. The effect of the spacetime 2-form
field on the worldvolume theory is far less clear.

Another interesting question is the relation between the field theory
we discuss, involving the 3-brane in the orientifold background, and
the field theory describing a 3-brane moving in the F-theory
background with no 2-form fields turned on. We have not identified
what the discrete torsion which differentiates between the two
theories corresponds to in the 3-brane field theory,
so the relation between these field theories is not clear. As
discussed in the next section, the discrete torsion definitely changes
the field theory at the origin of moduli space, although (since the
discrete torsion is concentrated at the origin) it does not change the
Coulomb branch away from the origin. Thus, the 3-brane moving in the
F-theory background should be described by a different field theory
(which should have exact $SL(2,\bZ)$ electric-magnetic duality), but
the Coulomb phase of all these theories should be identical. In the
F-theory background, tensionless strings appear in spacetime, and the
effect these should have on the 3-brane probe field theory is also
unknown.

We can easily generalize the analysis of this section to multiple
3-brane probes, as in section 2. We find for $k$ 3-brane probes an
$Sp(2k)\times Sp(2k)$ gauge theory, with two chiral superfields in the
$\bf(2k,2k)$ representation from $X_{6,7,8,9}$, superfields in the
$\bf(k(2k-1)-1,1)+(1,1)+(1,k(2k-1)-1,1)+(1,1)$ representations from
$X_{4,5}$, and eight $\bf(2k,1)$ and $\bf(1,2k)$ superfields from the
3-7 strings. The Coulomb phase of this theory may be easily seen to
consist of $k$ copies of the Coulomb phase of the original theory (up
to an $S_k$ identification), 
and we run into the same problems when trying to write a
superpotential describing the theory at the origin of the moduli space.

At other types of intersection points we also expect to find
$N=1$ superconformal theories with various global symmetries. For
instance, at the intersection of two $E_n$ singularities, we should
find an $N=1$ superconformal theory with $E_n\times E_n$ global
symmetry. As in \aks, we can compute the dimensions of some of the
operators in these $N=1$ SCFTs, but so far we do not have a lagrangian
that would flow to them, hence we cannot say much more about them.

\newsec{Summary and open questions}

In this paper we discussed two field theories arising from 3-brane
probes in F-theory. In section 2 we discussed the field theory
corresponding to $k$ parallel 3-brane probes in F-theory compactified
on $\bT^4/\bZ_2$. As expected from string theory, we found that the
Coulomb branch of the corresponding field theory was equivalent to $k$
copies of the Coulomb branch of the
original $k=1$ theory, divided by $S_k$. We expect this
to be the general result for multiple D-brane probes.

In section 3 we discussed a 3-brane probe at an intersection of two
$D_4$ singularities, in a particular orientifold vacuum of string
theory. By general considerations we expect this to correspond to an
$N=1$ superconformal field theory, which can flow (in two different
ways) to the Seiberg-Witten $N=2$ $SU(2)$ $N_f=4$ theory. The field
content we
found on the probe was consistent with this expectation, but we have
not been able to write down an explicit superpotential that would be
consistent with everything we know. Specifically, we were not able to
write down a superpotential that would give masses to some of the
quarks when we flow to the $N=2$ theory, and that would naturally
lead to the existence of a fixed line in the field-theory moduli
space. We are not sure about the meaning of this failure. 
It might be related to the interaction of the
probe with a tensor multiplet in spacetime, which may prevent us from
having a simple local description. It would be very interesting to
understand the resolution of this problem, which could also be related
to the problems encountered in the spacetime description of
intersecting singularities.

Next, we would like to discuss is the correspondence between
the spacetime fields and the parameters on the D-brane probe theory.
It is clear that any spacetime field should correspond to a
parameter in the worldvolume theory, since the D-brane couples in a
consistent way to the spacetime theory. However, a relation in the
opposite direction does not necessarily have to exist. In principle, 
there could
exist parameters of the worldvolume theory that do not correspond to
any spacetime fields, and the corresponding theories would not be
realized in the moduli space of a D-brane moving in the string theory
vacuum. One example of this phenomenon is the theory we discuss in
section 2, where a mass term for the antisymmetric $Sp(2k)$ field is
allowed in the field theory, but the corresponding parameter does not
exist in string theory (as noted also in \bhoo). 

Another example involves the quark masses in the theory we discuss in
section 3. In 8 dimensions, there was \refs{\sen,\bds} an exact
correspondence between the deformations of the spacetime theory and
the ($N=2$ preserving) parameters of the worldvolume theory. In
particular, the $SO(8)$ adjoint scalars in spacetime (which were part
of the vector multiplet), corresponding to moving the 7-branes off the
orientifold, were interpreted as a mass matrix for the
quarks in the worldvolume theory.
In the 6D generalization we considered in section 3 the situation is
different and more complicated, as we will now explain.
The orientifold we consider, corresponding to a compactification of
F-theory on $\bT^6 / \bZ_2 \times \bZ_2$,
is {\it not} connected in any simple way to the elliptic fibration over
${\bf F}_0$.
Instead, the orientifold compactification is connected
to a compactification of F-theory on a Calabi-Yau manifold with Hodge numbers
(51,3) \refs{\bz,\dabpark,\gm}. In this case
the orientifold analysis reveals that in 6D spacetime we do not have
any charged fields. Thus, we cannot break the $SO(8)$ gauge symmetry
by the appropriate Higgsing and, therefore,
we do not get quark mass parameters on the three-brane worldvolume
theory. Note that
at the orientifold point we have
altogether 17 tensor multiplets and 4 hypermultiplets. This field content
is consistent with the cancelation of 6D anomalies\foot{Note
that the disappearance of the $SO(8)$ adjoint field in the orientifold
procedure is consistent with the anomaly equation for $\tr(F^4)$, 
due to the relation
$\tr_{{\bf 28}}F^4=3 (\tr_{\bf 8}F^2)^2$ 
(without any $\tr_{\bf 8}F^4$ term).}.
The tensor multiplet localized at each
intersection point of groups of seven-branes
contains a scalar field, which provides a possible
deformation of the theory.
However, this deformation does not resolve the $D_4$ singularities,
but just separates the two intersecting singularities \bj. The four
hypermultiplets in spacetime correspond to moduli of the
compactification manifold, and do not affect our analysis near a
single orientifold point.
Thus, the moduli of the orientifold theory do not include fields
corresponding to quark mass matrices in this case.
To get such fields, we should go through phase transitions which would
turn the 16 extra tensor multiplets into hypermultiplets, taking us
to the general curve in Weierstrass
form \mv\ which has many (243) parameters. The modulus of the F-theory torus,
$\tau$, will, generically, depend
on all these parameters. Some of these parameters will presumably
correspond to quark mass matrices.
It is a challenging problem to understand this phase transition (which
is similar to the M-theory phase transition in which a fivebrane (tensor)
is ``swallowed'' by a 9-brane and turns into a large $E_8$ instanton)
and its effects on the 3-brane worldvolume field theory.

Finally, we now present a brief analysis
of the model of Sen \sentwo\ along the lines of section three.\foot{This 
is an extended version of a note added to our original 
preprint version. It was sparked by remarks of A.Sen, who also 
suggested that for his model there might be just the right field 
content on the probe to satisfy all the conditions we have required for
the superpotentials. This is what we are going to show here.}
But before doing that, we want to point out some of the differences
of the two models; for details we refer to \refs{\bz,\dabpark,\gm}. 
Both models can be described as type IIB on K3 orientifolds. They differ,
however, in the action of the orientifold projections on the twisted sector
and open string states.
The model considered by Sen \sentwo\ 
is T-dual to the Gimon-Polchinski model, and
thus does not contain any extra tensor multiplets, while the model we have
considered has extra tensor multiplets and is clearly different 
(it has been constructed explicitly in \refs{\bz,\dabpark,\gm}).
{}From the F-theory point of view, it seems that both models correspond
to the same elliptic curve \fibration-\phis, but they differ by
discrete 2-form fields concentrated at the intersection of the
singularities (as in the analogous string theory compactifications
discussed in \disctor). 
The two models both have such discrete torsion turned
on, so that none of them correspond to the naive F-theory
compactification, but the values of the discrete 2-form fields are
different in the two cases. In both cases the discrete 2-form fields
prevent the 3-brane from wrapping the collapsed 2-cycle and giving
rise to tensionless strings. In Sen's model the low-energy spectrum in
spacetime is also different from the naive F-theory spectrum,
suggesting that in this case both 2-form fields are turned on, while
in the orientifold we discussed in section 3 apparently only one of
the 2-form fields is turned on. As before, our computation here will
be in the orientifold theory, and its direct relation to F-theory is
not clear.

It is straightforward to repeat the analysis which leads
to the spectrum on the three-brane probe for the model considered by Sen. 
There is no change in our discussion of the states arising from the 
3-3 strings.
Since the model considered by Sen
is just the $T$ dual of the model studied in detail
by Gimon and Polchinski \gp\ -- the $T$ duality acting, say, in the $89$
plane -- we can immediately read off the matrices $\gamma$ acting on 
the sevenbrane indices of states in the 7-7, 
7-7', 7'-7' sectors and in the 3-7 and 3-7' sectors, from the ones
given in \gp, by $T$-dualizing. One finds that
$$
\gamma_{\Omega_1}^{(7)}=\gamma_{\Omega_2}^{(7')}={\bf 1}_{8\times 8}
$$
$$
\gamma_{T}^{(7)}=\gamma_{T}^{(7')}
=\gamma_{\Omega_2}^{(7)}=\gamma_{\Omega_1}^{(7')}
=\pmatrix{0&i{\bf 1}_{4\times 4}\cr -i{\bf 1}_{4\times 4}&0\cr}.
$$
Note that the $\gamma^{(7)}$ and $\gamma^{(7')}$ matrices are not all 
proportional to the identity matrix, reflecting the fact that the 
space-time gauge symmetry is no longer $SO(8)$, but in fact the subgroup
$U(4)$. The global symmetry of the probe theory is thus $U(4)\times U(4)$
instead of $SO(8)\times SO(8)$; see also \sentwo. This is part of the
$U(4)^8$ gauge symmetry of the spacetime theory \refs{\gp,\sentwo}.
Also, from \gp\ we 
learn that in the space-time theory there are, in addition to the 
gravity multiplet, the $U(4)\times U(4)$ gauge fields and one 
anti-selfdual tensor multiplet, also massless 
hypermultiplets: $2({\bf 6},{\bf1})+2({\bf1},{\bf6})+({\bf 4},{\bf
4})$ (which will correspond to mass terms in our probe theory).
Redoing our analysis of the 3-7 and 3-7' states with the new 7-brane
gamma matrices, we find states in the following representations of the
local and global symmetry group
$SU(2)\times SU(2)\times U(4)\times U(4)$:
$({\bf 2},{\bf 1},{\bf 4},{\bf 1})+({\bf 1},{\bf 2},\bar{\bf 4},{\bf 1})+
({\bf 2},{\bf 1},{\bf 1},{\bf 4})+({\bf 1},{\bf 2},{\bf 1},\bar{\bf 4})$. 
In addition to these chiral multiplets, 
we still have, as before,  the chiral superfields $A,B$ from the 3-3 
strings, both 
transforming as $({\bf 2},{\bf 2},{\bf 1},{\bf 1})$. 
Given the field content of this probe theory, we can now easily 
write down a superpotential which satisfies all the conditions described
in section 3. This probe theory thus provides the question to
which Sen \sentwo\ has given the string theory answer. Denoting
$({\bf 2},{\bf 1},{\bf 4},{\bf 1})\cor Q,\,
({\bf 1},{\bf 2},\bar{\bf 4},{\bf 1})\cor \tilde Q,\,
({\bf 2},{\bf 1},{\bf 1},{\bf 4})\cor q,\,
({\bf 1},{\bf 2},{\bf 1},\bar{\bf 4})\cor\tilde q$ 
one can now write a superpotential of the form $W = Q A \tilde Q + q B
\tilde q$ so that the conditions of ref.\ls\ 
for having a
fixed line passing through the origin of the space of couplings are
fulfilled.
This provides
an explicit realization of an $N=1$ supersymmetric theory which has an
exact electric-magnetic duality, as is evident from the string theory
answer. The duality group in this case, due to the presence of the
discrete torsion, is the $\Gamma(2)$ subgroup of $SL(2,\bZ)$.
Turning on an expectation value for $A$ ($B$) gives masses to the 
$Q,\tilde Q$ ($q,\tilde q$) quarks, and the
theory flows to the $N=2$  $SU(2)_D$ theory of \sen. 
Recall that $B$ transforms under the unbroken $SU(2)_D$ as 
${\bf 3}+{\bf 1}$ so that 
a superpotential of the form $qX\tilde q$, where $X$ is in the adjoint
of $SU(2)_D$,  emerges form the $qB\tilde q$ term.

We are tempted to conjecture that the difficulties that we encountered in
our original model, which are absent in the model considered by Sen, are
related to the fact that in the former we have extra tensor multiplets
which are absent in the latter. 
As discussed above, the two models seem to differ only by discrete
2-form fields concentrated at the singularities. Since these 2-form
fields can assume only discrete values, one cannot relate the two
theories simply by changing these fields continuously. It would be
interesting to understand if the theories could be related by some
sort of phase transition. Such a phase transition 
would obviously need to transform the
16 extra tensor multiplets in the first orientifold we analyzed into
hypermultiplets in the second orientifold (similar to the phase
transition related to the small $E_8$ instanton). This issue deserves
further investigation.

\bigskip

\centerline{\bf{Acknowledgements}}

We would like to thank P. Aspinwall, S. Kachru, P. Mayr, R. Plesser,
N. Seiberg, E. Silverstein and E. Witten 
for useful discussions. We thank, in
particular, V. Kaplunovsky for comments and for sharing with us his
insight on issues related to section 2. Special thanks also to 
A. Sen for communications after the first version of this paper
appeared. The work of OA was supported in part by DOE grant DE-FG02-96ER40559.

\vfill\eject

\appendix{A}{Explicit determination of the representations of the vector
and chiral superfields}

\subsec{Fields from the 3-brane strings}

First we determine the fields that
correspond to open strings with both ends on the 3-brane (or on its
images). In flat space the 3-brane field theory involves an $N=4$
vector multiplet, containing an $N=1$ vector multiplet and three $N=1$
chiral multiplets corresponding to the coordinates $X_{4,5,6,7,8,9}$
of the 3-brane. In the presence of the orientifold, each of these
fields is enhanced into a $4\times 4$ matrix with different
constraints.

\medskip
\centerline{(i) \bf Vector superfields}

 The relations imposed by \mishgam\ on the
components $M_{ij}$  of gauge fields are

\eqn\mishgcons{\eqalign{ M_{11}&= -M_{22} = -M_{33} = M_{44} \cr
 M_{14}&= M_{41} = M_{23} = M_{32} \cr
 M_{12}&= -M_{43} \quad M_{13} = -M_{42} \cr
M_{21}&= -M_{34} \quad M_{31} = -M_{24}. \cr}}

A basis of 6 matrices that obey these relations is
\eqn\vactormats{\eqalign{Z_1 = \pmatrix{ 1 & 0 & 0 & 0 \cr 0 & -1 & 0 & 0 \cr
0 & 0 & -1 & 0 \cr 0 & 0 & 0 & 1 \cr}  \quad &
Z_2 = \pmatrix{0 & 0 & 0 & -1 \cr 0 & 0 & -1 & 0 \cr 0 &
-1 & 0 & 0 \cr -1 & 0 & 0 & 0 \cr} \cr
W_1 = \pmatrix{ 0 & 1 & 0 & 0 \cr 1 & 0 & 0 & 0 \cr
0 & 0 & 0 & -1 \cr 0 & 0 & -1 & 0 \cr}  \quad &
W_2 = \pmatrix{0 & 0 & 1 & 0 \cr 0 & 0 & 0 & -1 \cr 1 &
0 & 0 & 0 \cr 0 & -1 & 0 & 0 \cr} \cr
W_3 = \pmatrix{ 0 & 0 & i & 0 \cr 0 & 0 & 0 & i \cr
-i & 0 & 0 & 0 \cr 0 & -i & 0 & 0 \cr}  \quad &
W_4 = \pmatrix{0 & i & 0 & 0 \cr -i & 0 & 0 & 0 \cr 0 &
0 & 0 & i \cr 0 & 0 & -i & 0 \cr}.}}

It is now straightforward to check that the matrices
$$\eqalign{W^+_2=&{1\over 4}(W_1 + W_2);\ W^-_1={1\over 4}(W_1 -
W_2); \ W^+_1={1\over 4}(W_3 + W_4); \ W^-_2={1\over 4}(W_3 -
W_4) \cr Z^{\pm}=&{1\over 4}(Z_1\pm Z_2)\cr}$$
obey the following commutation relations

\eqn\mishcr{\eqalign{ [Z^+, W^+_2]=&-i W^+_1;\ \  [Z^+, W^+_1]=i
W^+_2;\ \  [W^+_2, W^+_1]=-i Z^+\cr [Z^-, W^-_1]=&i W^-_2;\ \  [Z^-,
W^-_2]=-i W^-_1;\ \ [W^-_1, W^-_2]=i Z^-\cr [Z^+, W^-_1]=&0;\ \
[Z^+, W^-_2]=0;\ \ [Z^-, W^+_2]=0;\ \  [Z^-, W^+_1]=0;\ \
[Z^+,Z^-]=0\cr [W^+_2, W^-_1]=&0;\ \  [W^+_2, W^-_2]=0;\ \ [W^+_1,
W^-_1]=0;\ \  [W^+_1, W^-_2]=0. \cr
}}

Thus, we see that the gauge fields span an
$SU(2) \times SU(2)$ algebra.
When we take $z_1$ to infinity all 1-2,1-4,2-3 and 3-4 strings become
infinitely massive, so we can drop those wave-function matrices
with non-zero entries in those positions. This leaves
$Z_1,\,W_2$ and $W_3$, which generate an $SU(2)$ subalgebra.
Likewise, if we take $z_2$ to infinity all 1-3,1-4,2-3 and 2-4 strings
become massive, and we are left with
$Z_1,\,W_1$ and $W_4$, which generate a different $SU(2)$ subalgebra
(which does not commute with the previous one).
This  picture is in accord with the naive expectation following \sen\
of having just a single $SU(2)$ near $z_1=0$ or $z_2=0$.

\medskip
\centerline{(ii) \bf $X_{67}$ chiral multiplet}

Next consider the  implications of \mishgam\ on the scalar
fields   $X_{67}$.  The constraints on the
matrix components are now

\eqn\mishconsb{\eqalign{ M_{11}&= -M_{22} = M_{33} = -M_{44} \cr
 M_{14}&= -M_{41} = -M_{23} = M_{32} \cr
 M_{12}&= +M_{43} \quad  M_{13} = M_{42}=0 \cr
M_{21}&= +M_{34} \quad  M_{31} = M_{24}=0 \cr}}

A basis of hermitian matrices that obey the relations is the following
\eqn\xsixmats{\eqalign{A_1 = \pmatrix{ 1 & 0 & 0 & 0 \cr 0 & -1 & 0 & 0 \cr
0 & 0 & 1 & 0 \cr 0 & 0 & 0 & -1 \cr}  \quad &
A_2 = \pmatrix{0 & 0 & 0 & i \cr 0 & 0 & -i & 0 \cr 0 &
i & 0 & 0 \cr -i & 0 & 0 & 0 \cr} \cr
A_3 = \pmatrix{ 0 & 1 & 0 & 0 \cr 1 & 0 & 0 & 0 \cr
0 & 0 & 0 & 1 \cr 0 & 0 & 1 & 0 \cr}  \quad &
A_4 = \pmatrix{0 & i & 0 & 0 \cr -i & 0 & 0 & 0 \cr 0 &
0 & 0 & -i \cr 0 & 0 & i & 0 \cr}. \cr}}
Note that now if we take $z_1$ to infinity (naively) we are left only
with $A_1$, while if we take $z_2$ to infinity (naively) we remain
with $A_1,A_3$ and $A_4$ which are in the adjoint representation of
the remaining $SU(2)$, and thus we go over to the  picture of \sen,
as expected.

Define now the  matrices
$A_{-+},A_{++},A_{--}$ and $A_{+-}$ by the equations
$A_{++}=A_3-iA_4$;\  $A_{-+}=A_1+iA_2$;\ $A_{+-}=A_1-iA_2$ and
$A_{--}=A_3+iA_4$.  It can easily be checked that
$A= (A_{++},A_{-+},A_{+-},A_{--})$ are in the
$\bf(2,2)$ representation of $SU(2) \times SU(2)$, with the subscripts
corresponding to the charges under $Z^+$ and $Z^-$ (note that these
matrices are not hermitian). For instance, we
have

$$ [Z^+,A_{++}] = \ha A_{++} \ \ [Z^+,A_{-+}] = -\ha A_{-+} $$
$$ [W^+_2,A_{++}] = -\ha A_{-+} \ \ [W^+_2,A_{-+}] = -\ha A_{++} $$
$$ [W^+_1,A_{++}] = {i\over 2} A_{-+} \ \ [W^+_1,A_{-+}] = -{i\over 2}
A_{++} $$
and so on.

\medskip
\centerline{(iii) \bf $X_{89}$ chiral multiplet}

For the relations on   the $X_{89}$ chiral superfield
we would find similar (but not identical)
results to those of the previous section,
with the second and third rows and columns of all
matrices interchanged.
Thus, this chiral superfield, that we denote by
$B= (B_{++},B_{-+},B_{+-},B_{--})$,
is also
in the $\bf(2,2)$ representation of $SU(2) \times SU(2)$ (though
they are not represented in the same way).

As a consistency check, we verify
that a VEV for $X_{67}$, for instance, indeed breaks the gauge
symmetry to a diagonal $SU(2)$. For instance, taking $A_1$ to have
a VEV would leave exactly the matrices $Z_1,W_2$ and $W_3$ (given
above), which commute with it, as expected. The $A$'s would also all
become massive except $A_1$ (since they do not commute with it),
again as expected (since after moving along the flat direction we
should have just a single scalar). VEVs for both $X_{67}$ and
$X_{89}$ can also be analyzed and they all behave as expected from
naive considerations.

\medskip
\centerline{(iv) \bf $X_{45}$ chiral multiplet}

The relations for $X_{45}$ are
\eqn\mishconsb{\eqalign{ M_{11}&= M_{22} = M_{33} = M_{44} \cr
 M_{14}&= M_{41} = -M_{23} = -M_{32}, \cr}}
and their solutions are spanned by the two singlets $S_1$ and $S_2$
\eqn\xfourmats{S_1 = \pmatrix{ 1 & 0 & 0 & 0 \cr 0 & 1 & 0 & 0 \cr
0 & 0 & 1 & 0 \cr 0 & 0 & 0 & 1 \cr}  \quad
S_2 = \pmatrix{0 & 0 & 0 & 1 \cr 0 & 0 & -1 & 0 \cr 0 &
-1 & 0 & 0 \cr 1 & 0 & 0 & 0 \cr}.}

\subsec{Fields form the strings between the 3-brane and 7-branes }

As was mentioned in section 3 the
 $\gamma_{\Omega}$ matrices for the
7-branes  are
all proportional to the identity matrix \refs{\bz,\dabpark,\gm}.
The constraint on the
7-brane gauge fields is then just $M = - M^T$, giving an anti-symmetric
matrix corresponding to an $SO(8)$ space-time
gauge symmetry, since there are 8 7-branes
when including the $Z_2$ partners.

Next, we should use these matrices to analyze the wave function
of the 3-7 strings (as in \gp). The 7-brane side is
trivial, so the orbifold/orientifold projections
just mix the various 3-branes
according to the $\gamma_{\Omega}$ matrices  given in \mishgam;
for instance  taking
the $3_1-7$ state (where $3_1$ is the first 3-brane) via
$\gamma_{\Omega_1}$ to $i$ times
the $7-3_2$ state (with opposite orientation !),
via $\gamma_{\Omega_2}$ to $i$ times the $7-3_3$
state, and via $\gamma_{T}$ to minus the $3_4-7$ state.

A basis for the states going to a
specific 7-brane can thus be chosen to be
\eqn\quarks{D^+_u=(1,0,0,-1); \qquad D^+_d (0,1,1,0).}
Using the matrices we found above for the gauge fields,
it is easy to check that these are doublets under $SU(2)_+$, and
singlets with respect to
$SU(2)_-$. The corresponding chiral superfields are thus in
a $\bf(2,1)$ representation of
$SU(2)_+ \times SU(2)_-$.
{}For the other group of 7-branes we can then do
the same thing, but with minus the identity matrix for some of the
relevant 7-7 $\gamma_{\Omega}$ matrices (say for
$\gamma_{\Omega_2}$ and $\gamma_T$). Then, the basis
comes out to be
$$D^-_u=(1,0,0,1); \qquad D^-_d (0,1,-1,0). $$
which is in the $\bf(1,2)$
representation of  $SU(2)_+ \times SU(2)_-$.

\listrefs

\end